\documentclass[
    ,final            
  ,numberedheadings 
  ]
  {aipproc}

\layoutstyle{6x9}

\begin{document}

\title{``Black Star'' or Astrophysical Black Hole?}

\classification{04.70.-s, 04.70.Bw, 97.60.-s, 97.60.Lf}
\keywords      {black holes, general relativity}

\author{K. Petrovay}{
  address={E\"otv\"os University, Department of Astronomy, Budapest, Pf.~32,
  H-1518 Hungary\\
  and\\
Theoretical Institute for Advanced Research in Astrophysics, Dept. of Physics,\\ 
National Tsing Hua University, Hsinchu 30013, Taiwan
} 
}

\begin{abstract}
Recently wide publicity has been given to a claim by  T. Vachaspati that ``black
holes do not exist'', that the objects known as black holes in astrophysics
should rather be called ``black stars'' and they not only do not have event
horizons but actually can be the source of spectacular gamma ray bursts. In this
short essay (no flimsier than the original preprint where these extravagant
claims appeared) I demonstrate that these ill-considered claims are clearly
wrong. Yet they present a good occasion to reflect on some well known but little
discussed conceptual difficulties which arise when applying relativistic
terminology in an astrophysical context.
\end{abstract}

\maketitle


\section{A Question Every Physics Student Asks (and very few get an
answer...)}

I remember when, as an undergraduate physics student, I was first confronted
with the peculiar properties of black holes. Like many other students, I was
particularly bothered by the concept that while a falling object can reach the
event horizon in a finite proper time, the same will take infinitely long when
measured by an external observer (coordinate time). By the same argument,
the formation of the black hole in gravitational collapse would also take an
infinitely long time for the observer. So how can we talk about black holes
having been formed in the collapse of dying stars? 

When I asked my professor, he suggested that I should think about it this way:
the event horizon has already formed but, owing to the strong time dilatation,
we have no knowledge of it yet, nor will we until the end of time. Which,
incidentally, is the very reason why nothing beyond the horizon can ever be
seen, i.e. why the horizon is a horizon and why the black hole is a black hole. 

This reply is ultimately based on the intuitive notion that proper time is, in
some sense, indeed the ``proper'' time, i.e. that the time difference of two
events (such as the beginning of gravitational collapse and the formation of the
event horizon) should be measured in a frame where they occur at the same place.
Indeed, there is a time honoured tradition in astronomy to apply {\it light-time
correction\/} to observations when giving ephemerides. In this vein, talking
about astrophysical black holes follows from the application of the principle of
light-time correction, which in this case happens to be infinite.

While this kind of ``light-time argument'' is the most common way astronomers
deal with the above conceptual problem, there is clearly something
unsatisfactory about it. General relativity is based on the principle that the
laws of physics are the same in all frames of reference, so no frame is more
legitimate than the other. Yet the above argument is based on the assumption
that proper time is, somehow, the ``real'' time and coordinate time is just some
appearance. But then, it is exactly appearences we are dealing with in astronomy
---so if we make such an arbitrary distinction between the two frames, would it
not be more plausible to base our terminology on what is measured in coordinate
time?

While it is true that in the observer's frame every single particle of the
collapsing star will in principle remain observable until the end of time, this
is a purely hypothetical observability. The reality is that any radiation
emitted from collapsing star's material, suspended just above the Schwarzschild
radius, will be gravitationally redshifted$+$dilatated into oblivion. Thus, from
the observer's point of view, the object behaves just like an already formed
black hole: the collapsed material has disappeared from sight for good, while
the behavior of anything that is far enough from the would-be horizon to be
observable is governed by the Schwarzschild (or Kerr) metric.

The ultimate justification of talking about ``astrophysical black holes''
lies in this empirical indistinguishability of the objects from {\it bona
fide\/} black holes.

\section{An Extravagant Claim (and why it's obviously wrong...)}

It is this indistinguishability that is challenged by Vachaspati
\cite{Vachaspati:noBH} who claims that the collison of two ``astrophysical black
holes'' may result in readily observable ---indeed, spectacular--- effects. If
this were true, it would indeed invalidate the use of the term ``black hole''
for these astrophysical objets. Vachaspati proposes ``black star'' instead.

By order of magnitude estmates he shows that the energy released in the 
collision of two ``black stars'' is comparable to that of gamma ray bursts, and
it should be radiated in the high energy electromagnetic regime. If so, this
could be a possible mechanism to produce gamma ray bursts.

What he inexplicably seems to forget, however, is that

(a) The time it takes for two colliding ``black stars'', shrunken arbitrarily close
to the Schwarzschild radii, to actually get in contact is arbitrarily long in
coordinate time, for the same well known reasons as for a test particle falling
into the hole. Thus, currently observed gamma ray bursts (or any astronomical
phenomenon) cannot be attributed to this.

(b) The estimate of the emitted power is based on the collision timescale in
{\it proper time,} so even if such a gamma ray burst were produced in the
vicinity of the Schwarzshild radius, it would be gravitationally
redshifted$+$dilatated into oblivion, as any other kind of radiation.

\section{A Slightly Less Extravagant Other Claim (and why it's irrelevant to
astrophysics...)}

In a related earlier paper, Vachaspati, Stojkovic and Krauss \cite{Vachaspati:prehawking} study the radiation of
the material collapsing into a black hole in quantum field theory. They find
that the infalling material is fully radiated away in what they call
``pre-Hawking radiation'' before a black hole could form. If true, this would
mean that actually {\it no\/} black holes that were not present from the start
of time could {\it ever\/} form in {\it either\/} proper or coordinate time.
Assuming for a moment that this claim is not based on the same kind of fallacy
as the claim about gamma bursts, this claim would have far-reaching consequences
e.g. for the attempts to create black holes in colliders.

Nevertheless, even if it were proven correct, its consequences for astrophysics
would be rather limited. True, this would imply that a true black hole would not
form even in infinite coordinate time, so the light-time correction argument for
calling these objects black holes would fail. However, in macroscopic black
holes the quantum effects that give rise to Hawking and ``pre-Hawking''
radiation operate on time scales much longer than the age of the universe, so
the observed behavior of astrophysical black holes would not be influenced. And
we have already seen that the more fundamental reason for calling these objects
black holes is the impossibility of dinstinguishing them empirically from
already formed {\it bona fide\/} black holes. This principle is still alive and
well, implying that there is no need to drastically change our terminology. For
the pedantic, the ``astrophysical'' qualifier in front of ``black hole'' should
serve as sufficient reminder of the true nature of the objects we are dealing
with.


\begin{theacknowledgments}

This work was supported by the Theoretical Institute for Advanced Research in
Astrophysics (TIARA) operated under Academia Sinica and the National Science
Council Excellence Projects program in Taiwan administered through grant number
NSC95-2752-M-007-006-PAE, as well as by the Hungarian Science Research Fund
(OTKA) under grant no.\ K67746.

\end{theacknowledgments}



\end{document}